\documentclass[conference]{IEEEtran}
\IEEEoverridecommandlockouts

\usepackage{cite}
\usepackage{amsmath,amssymb,amsfonts}
\usepackage{algorithmic}
\usepackage{graphicx}
\usepackage{textcomp}
\usepackage{xcolor}
\usepackage{tikz}
\usepackage{array}
\usepackage{booktabs}
\usetikzlibrary{arrows.meta, positioning, fit, calc, backgrounds}
\usepackage[colorlinks=true, linkcolor=blue, citecolor=blue, urlcolor=blue]{hyperref}
\def\BibTeX{{\rm B\kern-.05em{\sc i\kern-.025em b}\kern-.08em
    T\kern-.1667em\lower.7ex\hbox{E}\kern-.125emX}}
\begin{document}

\title{Not What You Asked For: Typographic Attacks in Household Robot Manipulation\\
}

\author{
\IEEEauthorblockN{1\textsuperscript{st} Ali Iranmanesh}
\IEEEauthorblockA{\textit{Cyber Security Lab} \\
\textit{The Pennsylvania State University}\\
State College, USA \\
afi5220@psu.edu}
\and
\IEEEauthorblockN{2\textsuperscript{nd} Peng Liu}
\IEEEauthorblockA{\textit{Cyber Security Lab} \\
\textit{The Pennsylvania State University}\\
State College, USA \\
pxl20@psu.edu}
}

\maketitle

\begin{abstract}
Open-vocabulary embodied AI agents increasingly rely on
vision-language models such as CLIP for object perception
and task grounding. However, the shared embedding space that
enables this flexibility introduces a structural vulnerability
to typographic attacks, where printed text in a physical scene
semantically overrides visual judgment. While prior work has
quantified this threat in static 2D benchmarks and 3D
navigation tasks, its impact on the full Sense-Plan-Act
pipeline of household robot manipulation remains unexplored.

This work evaluates typographic attacks in a Habitat-based
simulation using the HomeRobot benchmark. We introduce a
decoupled perception architecture that exposes a frozen CLIP
encoder to adversarial stickers while maintaining geometric
grounding via DETIC. In a controlled evaluation pool of 59 attributable episodes, 
the attack achieves an overall Attack Success Rate (ASR) of 67.8\%, 
rising to 70.0\% among fully successful episodes, under uncontrolled 
viewing angles and occlusion with no perceptual optimization.

Critically, we find that perceptual errors propagate through the persistent 3D 
semantic map to produce kinetic failures, defined here as physically executed 
grasping and transport of the wrong object driven by an adversarially 
poisoned semantic state. In these cases, the robot physically grasps 
and delivers the wrong object to a target receptacle. These results establish
typographic misclassification as a real, measurable, and
physically consequential threat to the safety of modular
manipulation pipelines that prior typographic attack research
has left unexamined.
\end{abstract}

\section{Introduction}

The integration of large-scale vision-language models has
transformed household robotics from closed-set systems into
open-vocabulary agents capable of following natural language
instructions. Models like Contrastive Language-Image
Pretraining (CLIP)~\cite{clip2021} allow robots to identify
and interact with thousands of object categories without
task-specific retraining. This capability is foundational to
modern modular manipulation stacks, which combine
open-vocabulary perception with 3D semantic mapping and
heuristic planning to execute complex pick-and-place tasks in
unstructured residential environments.

However, the alignment of visual and linguistic features in a
shared embedding space creates a structural security risk
known as a typographic attack. By introducing printed text
into the physical environment, an adversary can cause a
vision-language model to misclassify an object as whatever
the text says, effectively overriding its visual
judgment~\cite{goh2021}. While this phenomenon is well
documented in 2D image classification, the implications for
embodied agents are significantly more severe. In a robotic
manipulation pipeline, a misclassification is not a momentary
error. It is written into a persistent 3D semantic map and
propagated through all downstream planning cycles. This
produces a kinetic failure where the agent physically commits
to the wrong object, grasps it, and delivers it to the target
receptacle.

Despite the rapid expansion of adversarial research in
embodied AI, household robot manipulation remains unexamined
in the context of typographic threats. SCAM~\cite{scam2025}
evaluated 99 vision-language models under real-world
handwritten typographic attacks and found an average accuracy
drop of 26 percentage points. CHAI~\cite{chai2026} achieved
attack success rates above 87\% against navigation agents
including autonomous drones and ground vehicles. Both
benchmarks, however, target either static image classification
or high-level navigation decisions. Neither captures the
failure modes specific to manipulation agents, where errors
involve distinguishing among small, cluttered household
objects and executing a physical grasp that irreversibly
alters the state of the world within the episode.

In this paper, we fill this gap by evaluating typographic
attacks against a full-stack HomeRobot~\cite{homerobot2023}
agent within the Habitat~\cite{habitat2019} simulation
environment. We introduce adversarial stickers labeled with
goal-object names into residential scenes to intercept the
agent's perception loop. Additionally, we develop a modified
OvmmPerception module that exposes CLIP to typographic cues
while preserving the geometric accuracy required for
navigation and manipulation.

Our contributions are as follows:

\begin{itemize}
    \item We identify and formalize household robot
    manipulation as an unexamined attack surface for
    typographic attacks. We characterize the kinetic failure
    mode that distinguishes it from prior evaluations: a
    perceptual misclassification that propagates through the
    full Sense-Plan-Act loop to produce a physical grasping
    error.

    \item We instantiate a \textbf{representative reference architecture} 
    for open-vocabulary manipulation that mirrors the joint-embedding 
    paradigm used in deployed systems like EmbCLIP and TidyBot. 
    While we independently validated the typographic vulnerability 
    against the third-party EmbCLIP agent, that system is restricted to 
    navigation, and our architecture isolates the vision-language 
    alignment stage to enable the first controlled evaluation of how 
    these confirmed perceptual flaws propagate through a full-stack 
    manipulation pipeline.

    \item We evaluate the attack across 59 attributable episodes 
    from a 1,199-episode validation set, the complete subset of episodes 
    in which failures are unambiguously attributable to the adversarial 
    perturbation rather than pre-existing agent limitations. Our results 
    show a 67.8\% overall ASR and a 70.0\% ASR on fully successful episodes, 
    achieved without perceptual optimization under uncontrolled physical conditions.
    \item We characterize the attack mechanism and phase-level
    consequences, showing that typographic cues are written
    into persistent spatial memory. This causes task failure
    even after the sticker leaves the agent's field of view,
    frequently resulting in the physical delivery of the wrong
    object to the receptacle.
\end{itemize}

By establishing that typographic misclassification is a
physically consequential threat in household robot
manipulation, this work highlights the need for robust
perceptual defenses in the next generation of
vision-language-action models.

\section{Background}\label{sec:background}

\subsection{Contrastive Language-Image Pretraining (CLIP)}

Object recognition in computer vision has historically depended on
fixed, closed-set vocabularies: a model trained to classify among
1{,}000 ImageNet categories cannot generalize to a novel label
without retraining. A foundational step toward breaking this
constraint was the construction of large-scale image-text corpora
that ground visual content in natural language descriptions rather
than discrete category indices~\cite{visualgenome2017}. Building
on this direction, Radford et al.~\cite{clip2021} introduced CLIP
(Contrastive Language-Image Pretraining), a dual-encoder model
trained on 400 million image-text pairs scraped from the web. CLIP
trains a {\bf vision encoder} and a text encoder jointly by maximizing
the cosine similarity of matching image-text pairs and minimizing
it for non-matching pairs within each batch, forming a contrastive
objective that aligns visual and linguistic representations in a
shared embedding space. At inference, classification is performed
without any task-specific training: a text prompt such as
\textit{``a photo of a \{class\}''} is encoded alongside the image,
and the class whose text embedding scores highest against the image
embedding is selected. This zero-shot transfer capability made CLIP
the perceptual backbone of choice for open-vocabulary embodied AI agents
such as EmbCLIP~\cite{embclip2022}, TidyBot~\cite{tidybot2023},
and CLIPort~\cite{cliport2021}. Subsequent vision-language-action (VLA) models including
OpenVLA~\cite{openvla2024} transitioned to SigLIP~\cite{siglip2023}
as their vision encoder, which replaces CLIP's softmax contrastive
loss with a sigmoid-based objective yet remains vulnerable to
typographic attacks~\cite{scam2025}, confirming that the threat extends
across the joint vision-language embedding paradigm.

The same embedding alignment that enables open-vocabulary generalization, 
however, introduces a structural vulnerability.
Because CLIP's text encoder and vision encoder share a joint
embedding space, any text visible within the image can activate
text-like features in the vision encoder and {\bf compete} with purely
visual evidence for the final similarity score~\cite{dyslexify2025}.
In embodied settings where the agent receives crop-level RGB inputs
for each detected instance, this competition occurs at the scale of
individual bounding-box crops, making the classifier sensitive to
text that appears anywhere within the crop's field of view.

\subsection{Typographic Attacks}

Typographic attacks exploit CLIP's joint embedding space by
introducing printed text into the physical scene that semantically
overrides the model's visual judgment. The phenomenon was first
characterized by Goh et al.~\cite{goh2021}, who demonstrated that
affixing a handwritten label to an object causes CLIP to classify
the object as whatever the label says, regardless of its visual
appearance, a behavior they described as the model
``reading first, looking later.''~\cite{goh2021} Mechanistic work
has traced this sensitivity to specialized attention heads in the
later layers of CLIP's vision encoder, where typographic tokens
are transferred to the class token before the final similarity
computation~\cite{dyslexify2025}; the vulnerability is structurally
encoded in the attention circuit and persists across model sizes.

The practical severity of this vulnerability at scale has been
quantified by the SCAM benchmark~\cite{scam2025}, the largest
real-world typographic attack dataset to date, which evaluated
99 vision-language models and found an average accuracy drop of
26\% under typographic attack. SCAM also
establishes that synthetic attacks closely mirror real-world
handwritten ones, directly motivating the use of physical stickers
as the attack medium in our evaluation. SceneTAP~\cite{scenetap2024}
extended the attack surface further by using an LLM-driven planner
to generate scene-coherent adversarial text placed at locations
most likely to deceive the model while remaining visually natural
to human observers, and successfully misled state-of-the-art
vision-language models even when deployed in physical environments.

\subsection{Household Robot Manipulation}

Household robots perform pick-and-place tasks in unstructured
indoor environments, navigating to a goal object specified by
a free-form text prompt, grasping it, and delivering it to a
target receptacle. A single execution of this task is called
an episode, which instantiates a specific scene, goal object,
and agent starting pose. Modular manipulation agents typically combine an open-vocabulary
object detector, a persistent 3D semantic map, and a finite-state
task planner to execute these tasks. Evaluating attacks in this
setting requires a simulation platform that supports open-vocabulary
perception, 3D semantic mapping, and physical manipulation in
residential scenes. The platform used in this work and how
CLIP is introduced as the attack-exposed classifier within it
are described in Section~\ref{sec:environment}.

\section{Related Work}

\subsection{Embodied AI Architectures and VLA Models}

Early domestic robotics platforms follow a modular Sense-Plan-Act paradigm
in which perception, mapping, planning, and execution are distinct
components~\cite{homerobot2023}, introducing integration boundaries that
silently amplify perceptual errors through the pipeline. The field has moved
rapidly toward unified VLA models that collapse this stack into a single
end-to-end network~\cite{rt2_2023, openvla2024}. These architectures inherit
the semantic richness of large-scale vision-language pretraining but also
inherit its vulnerabilities: adversarial perturbations that previously
affected only classification can now directly influence motor outputs,
collapsing the safety margins that modularity once provided.

\subsection{Typographic and Adversarial Attacks on Robotic Stacks}

Typographic attacks on CLIP were first characterized by Goh et
al.~\cite{goh2021} and quantified at scale by SCAM, which
found an average accuracy drop of 26 percentage points across 99
vision-language models. SceneTAP~\cite{scenetap2024} extended this to
physical environments using LLM-driven scene-coherent adversarial text.
The broader VLA attack landscape reveals additional threat dimensions: adversarial
patches can reduce task success to zero~\cite{vlaattack2025},
Tex3D~\cite{tex3d2025} achieves failure rates of up to 96.7\% via optimized
3D textures, and FreezeVLA~\cite{freezevla2025} introduces action-freezing
attacks that render agents persistently unresponsive. CHAI~\cite{chai2026}
extends typographic attacks into fully embodied 3D settings by jointly optimizing
the semantic content and perceptual appearance of injected text signs against
Large Vision-Language Model (LVLM)-driven navigation agents, including autonomous drones and ground vehicles,
achieving attack success rates above 87\% in real-world robotic experiments.
On the supply-chain side, TrojanRobot~\cite{trojanrobot2024} and
LoRATK~\cite{loratk2024} demonstrate backdoor injection through modular policy
components and LoRA adapters respectively.

\subsection{Defenses}

Defense-Prefix~\cite{defenseprefix2023} mitigates typographic attacks by
prepending a learned token to each class name in CLIP's text encoder,
hardening the prompt against typographic interference without modifying model
parameters. Dyslexify~\cite{dyslexify2025} takes a more mechanistic approach,
identifying and ablating the small set of attention heads in CLIP's vision
encoder that causally transmit typographic information to the class token,
producing a model that resists typographic attacks without any retraining.
Both methods reduce but do not eliminate typographic misclassifications.
Fully eliminating the vulnerability at the encoder level in CLIP and its
successors such as SigLIP remains an open problem.

\section{Motivation: The Manipulation Domain Gap}

Typographic attack research has expanded significantly in scope, from static 2D benchmarks such as SCAM~\cite{scam2025} and SceneTAP~\cite{scenetap2024} that measure misclassification rates on fixed image datasets, to embodied 3D settings: CHAI~\cite{chai2026} evaluates physical-world typographic attacks against LVLM-driven navigation agents including autonomous drones and ground vehicles, and Tex3D~\cite{tex3d2025} targets VLA models through adversarial 3D object textures. Despite this breadth, one setting has received no coverage: household robot manipulation, where the agent must navigate a furnished indoor environment, identify and grasp a goal object from among household distractors, and place it at a target receptacle. This gap exists for three reasons that are specific to the manipulation domain.

\textbf{Reason 1: Navigation and driving agents measure decision-level outputs;In contrast, manipulation agents expose a distinct, propagating failure mode.}
CHAI and related 3D attack works measure whether an agent issues a wrong high-level command, such as landing on an unsafe rooftop or proceeding through a crosswalk. In a household robot manipulation pipeline, the failure is different in kind. A typographic misclassification at the object-classification stage is written into a persistent semantic voxel map and propagated with opaque persistence through all downstream
Sense-Plan-Act cycles. The agent navigates to the wrong target, attempts to grasp it, and physically picks it up, committing a kinetic error that is not a decision about where to go but about what object now exists in the robot's gripper. No prior 3D attack evaluation tracks this class of failure.

\textbf{Reason 2: Household object classification involves dense, cluttered scenes that are structurally different from navigation targets.}
Navigation and driving attacks operate in scenes where the adversarial target is a landmark, a rooftop, or a vehicle; these are spatially prominent and typically occupy a large portion of the agent's field of view. Household robot manipulation requires distinguishing among many small, visually similar household objects, such as cups, bottles, and containers, that share category-level features. Adversarial stickers placed on one distractor must compete against multiple plausible candidates in the same crop-level view passed to CLIP, making the attack surface and the conditions for misclassification accumulation structurally distinct from anything evaluated in prior 3D work.

\textbf{Reason 3: The pick-and-place outcome requires a manipulation platform that no existing 3D attack benchmark provides.}
CHAI evaluates against navigation and flight agents. Tex3D targets end-to-end VLA models but does not evaluate the full Sense-Plan-Act loop of a modular manipulation agent in a residential setting. The kinetic failure mode central to this work, where a misclassified household object is physically grasped and carried to a receptacle, is only observable in a simulator that couples open-vocabulary perception, 3D semantic mapping, heuristic planning, and snap-based manipulation in a household indoor scene. No existing attack benchmark, 2D or 3D, provides this infrastructure.

\textbf{Goal of this work.} We fill this gap by introducing typographic stickers into 
a Habitat-based household manipulation simulation and evaluating their 
effect on a full-stack HomeRobot agent across the complete Sense-Plan-Act loop.
 Our goal is not to improve attack strength or propose a new attack method, 
 but to establish that typographic misclassification is a real, measurable, 
 and physically consequential threat in the household robot manipulation 
 setting that the full body of prior typographic attack work, spanning 
 static 2D benchmarks, physical-world VLM evaluations, and 3D navigation 
 agents, 
has uniformly left unexamined.

\section{Threat Model}\label{sec:threat}

We consider an adversary who seeks to cause a household
manipulation agent to physically pick and deliver the wrong
object by placing typographic stickers in the agent's
operating environment. The adversary operates under the
following capabilities and constraints.

\textbf{Adversary capabilities.} The adversary has physical
access to the household environment and can place printed
stickers on flat surfaces along plausible navigation routes,
such as tables, chairs, and floors. The sticker bears the
goal object label $g_\tau$ in printed text. No cyber access
to the agent, its sensors, or its software stack is required.
The attack is entirely physical and requires no modification
to the agent's hardware or code.

\textbf{Adversary knowledge.} The adversary requires only class-level knowledge of the robot's operating domain 
(e.g., common targets like ``cups'' or ``bottles'') rather than episode-specific prompts. 
A single printed label bearing the goal object name is sufficient to intercept the 
agent's perception pipeline. This models a low-resource physical adversary such as 
an insider or a visitor with brief access to the space.

\textbf{Attack surface.} The attack targets the
threshold-gated CLIP override mechanism in the OvmmPerception
module. When a sticker bearing $g_\tau$ enters a DETIC
bounding box crop and drives the cosine similarity score
above $\theta$, CLIP overrides DETIC's semantic label for
that instance. The pipeline contains no input sanitization,
anomaly detection, or label provenance tracking, making the
override fully exploitable through crafted visual text alone.

\textbf{Persistence via semantic map poisoning.} Once the
adversarial label is written into the 3D semantic voxel map,
it persists across all subsequent timesteps regardless of
whether the sticker remains in the agent's field of view.
The voxel map functions as a spatial cache: each voxel at
coordinates $(x, y, z)$ stores a semantic label that is
written on CLIP override and read by the state machine and
FMM planner on every subsequent cycle. A successful
typographic override constitutes a cache poisoning attack,
where a single malicious write at the perception stage
corrupts the agent's spatial memory and drives all
downstream navigation and manipulation decisions toward the
wrong object for the remainder of the episode. No defensive
mechanism intervenes between the poisoned cache entry and
the kinetic outcome.

\section{Household Robot Manipulation Environments}\label{sec:environment}

A household robot manipulation task requires an agent to navigate
a furnished indoor environment, identify a goal object specified
by a free-form natural language prompt, grasp it, and place it
at a designated receptacle. Each instantiation of this task is
called an episode: a unique combination of a residential scene,
a goal object, and an agent starting pose. Episodes vary across
room layouts, object arrangements, and goal categories, making
no two evaluation conditions identical.

What distinguishes this setting from navigation-only embodied
AI tasks is the combination of dense object clutter, persistent
3D semantic state, and a physical grasping outcome. Navigation
agents decide where to go; manipulation agents decide what
object to pick up and physically commit to that decision by
grasping it. Household scenes contain many small, visually
similar objects such as cups, bottles, and containers that
share category-level features and occupy the same cluttered
surfaces. A perception error in this setting is not a wrong
turn but a wrong object in the robot's gripper, carried
through the rest of the task.

This physical commitment creates a distinct challenge for
perception-level attacks. A typographic misclassification
does not merely alter a navigation decision; it is written
into a persistent 3D semantic map and propagated silently
through every downstream planning and manipulation step
until the task terminates with the wrong object physically
placed at the receptacle. No defensive mechanism intervenes
between the perceptual error and the kinetic outcome.

Habitat~\cite{habitat2019} is the simulation platform used
in this work, providing photorealistic residential scenes with
RGB-D observations, depth-grounded physics, and support for
pick-and-place tasks. HomeRobot~\cite{homerobot2023} is the
open-vocabulary mobile manipulation benchmark built on Habitat
that serves as our experimental platform, providing a pretrained
Sense-Plan-Act agent capable of completing the full manipulation
loop. The architecture of our modified agent and how CLIP is
introduced as the attack-exposed classifier are described in
Section~\ref{sec:approach}.

\section{Approach}\label{sec:approach}

\subsection{Technical Challenges}

\textit{1) Physical deployment and agent training from scratch are jointly infeasible for attack evaluation.}
Deploying a physical manipulation platform across multiple furnished evaluation rooms carries financial and spatial overhead that exceeds typical academic resources. Training a manipulation agent from scratch solely for attack evaluation is similarly prohibitive, and would yield a model of unverified task competence whose failures could not be cleanly attributed to the adversarial perturbation.

\textit{2) The pretrained agent succeeds on only a limited subset of available episodes.}
The HomeRobot validation set contains 1,199 episodes spanning diverse scenes and goal categories, but the pretrained agent completes the full pick-and-place loop in only a small fraction of them. Evaluating attacks across the full set would conflate pre-existing agent failures with attack-induced ones, making it impossible to isolate the typographic perturbation as the cause of any observed deviation.

\textit{3) The HomeRobot agent relies on DETIC for object detection, which does not expose CLIP to typographic attacks.}
In the unmodified HomeRobot stack, DETIC handles both spatial proposal generation and semantic labeling. Because DETIC does not use a joint vision-language embedding, it is not susceptible to typographic cues. Evaluating typographic attacks against this architecture would produce a null result by construction, independent of whether the vulnerability exists in the broader class of open-vocabulary agents we aim to model.

\subsection{Addressing Challenge 1}

An intuitive approach would be to deploy a physical robot in a household environment
and observe its behavior when adversarial stickers are introduced. This is infeasible
for two reasons. First, the financial and spatial overhead of obtaining a manipulation
platform and configuring multiple furnished evaluation rooms exceeds typical academic
resources; this is precisely the motivation behind sim-to-real frameworks such as
Habitat~\cite{habitat2019} and RoboTHOR~\cite{robothor2020}.
Second, training an embodied agent from scratch solely for attack evaluation is
prohibitively time-consuming and would produce a model whose task competence
is unverified. We therefore build on HomeRobot~\cite{homerobot2023} within Habitat, which
provides a pretrained agent that already succeeds at pick-and-place navigation,
allowing us to introduce typographic attacks directly against a capable agent
without incurring the overhead of environment construction or agent training.

\subsection{Addressing Challenge 2}

A naive approach would be to evaluate typographic attacks across all 1,199 episodes
in the HomeRobot validation set. This is problematic because any observed failure
must be attributable solely to the adversarial sticker, not to pre-existing
limitations of the agent. Evaluating on episodes where the agent already fails
without any attack present would introduce confounds that undermine the validity
of the results. We therefore restrict evaluation to the subset of episodes where
the unattacked agent either succeeds fully or fails only at the final placement
step, ensuring that any deviation from pre-attack behavior under attack can be
cleanly attributed to the typographic perturbation. Out of the full 1,199
episodes, the pretrained agent succeeds completely in only 10 and partially
in 49, leaving a constrained but unambiguous pool of valid evaluation candidates.
The full episode characterization and formal definitions are presented in
Section~\ref{sec:results} as part of RQ1.

\subsection{Addressing Challenge 3}

Since our work targets typographic attacks against CLIP specifically, retaining
Detecting Twenty-thousand Classes using Image-level
Supervision (DETIC)~\cite{detic2022} as the sole classifier would not expose the vulnerability under
investigation. Fully replacing DETIC with CLIP is equally problematic: CLIP is a
vision-language encoder that produces scalar text-image similarity scores and is
not designed to generate spatial proposals or segmentation masks, so a full swap
would leave the navigation stack without the geometric grounding it requires. We
therefore introduce CLIP as a parallel goal-scoring layer inside the OvmmPerception
module, running alongside DETIC, which is kept for spatial proposal generation
and mask prediction. This decoupled design makes CLIP solely responsible for the
final object classification decision and directly exposes it
to adversarial typographic stickers placed in the scene. It
is important to note that the unmodified HomeRobot agent
relies solely on DETIC, which is not vulnerable to typographic
cues. We intentionally introduce the CLIP-based override to
represent the architecture of modern open-vocabulary VLA
agents such as EmbCLIP~\cite{embclip2022} and SigLIP-based
systems~\cite{openvla2024}, which rely on joint vision-language
embeddings for object classification. HomeRobot was selected
as the evaluation platform because it is the only open-source,
simulation-available benchmark designed for open-vocabulary
mobile manipulation; alternative agents using CLIP or SigLIP
are either restricted to object navigation, closed-source, or
unavailable in simulation, making controlled full-stack
evaluation infeasible on those platforms.
We acknowledge that the unmodified HomeRobot agent uses DETIC alone and is not
natively vulnerable to typographic attacks. Introducing CLIP as a parallel
classification layer is an intentional architectural choice whose
representativeness warrants direct justification. To ensure external 
validity, we first confirmed the typographic vulnerability on the 
third-party EmbCLIP agent, achieving successful misclassification 
in its native object navigation tasks. While EmbCLIP is restricted 
to navigation, we utilize our modified HomeRobot stack as a 
\textbf{functional proxy} to evaluate how this confirmed vulnerability 
propagates through a full manipulation loop. Our decoupled architecture 
is an intentional choice to model the classification patterns of deployed 
open-vocabulary systems in a controllable form that allows us to measure 
downstream kinetic failures across the full Sense-Plan-Act loop.

\begin{figure*}[htbp]
\centering
\resizebox{\textwidth}{!}{%
\begin{tikzpicture}[
    >=Stealth,
    line width=1.2pt,
    font=\sffamily,
]

\tikzset{
  deticbox/.style    = {draw, fill=green!20, rounded corners=4pt,
                         minimum width=2.2cm, minimum height=1.3cm,
                         align=center, font=\small},
  clipbox/.style     = {draw, fill=blue!15, rounded corners=4pt,
                         minimum width=2.2cm, minimum height=1.4cm,
                         align=center, font=\small},
  overridebox/.style = {draw=red!80, very thick, fill=red!12, rounded corners=4pt,
                         minimum width=2.0cm, minimum height=1.3cm,
                         align=center, font=\small},
  voxelbox/.style    = {draw, fill=yellow!30, rounded corners=4pt,
                         minimum width=2.0cm, minimum height=1.3cm,
                         align=center, font=\small},
  smbox/.style       = {draw, fill=cyan!20, rounded corners=4pt,
                         minimum width=2.8cm, minimum height=2.4cm,
                         align=center, font=\small},
  envbox/.style      = {draw, fill=gray!15, rounded corners=4pt,
                         minimum width=2.6cm, minimum height=1.0cm,
                         align=center, font=\small},
  actbox/.style      = {draw, fill=orange!40, rounded corners=4pt,
                         minimum width=1.6cm, minimum height=0.9cm,
                         align=center, font=\small},
  wsbox/.style       = {draw, fill=white, rounded corners=4pt,
                         minimum width=1.6cm, minimum height=0.72cm,
                         align=center, font=\small},
  promptbox/.style   = {draw=purple!70, fill=purple!10, rounded corners=4pt,
                         minimum width=2.0cm, minimum height=0.9cm,
                         align=center, font=\small},
  lbl/.style         = {font=\scriptsize\sffamily},
}

\node (img) at (0, 1.6) {
    \includegraphics[width=2.9cm]{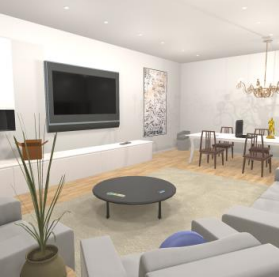}
};
\node[wsbox] (ws) at (0, -0.7) {World State};
\node[draw=black!60, dashed, rounded corners=3pt, inner sep=10pt,
      fit=(img)(ws)] (dbox) {};

\node[deticbox] (detic) at (4.5, 2.4) {
    \textbf{DETIC}\\Proposals \& Masks
};

\node[clipbox] (clip) at (4.5, 0.1) {
    \textbf{CLIP}\\(Goal Scoring)
};

\node[promptbox] (prompt) at (4.5, -1.9) {
    a photo of a\\\{goal\}
};

\node[overridebox] (override) at (8.2, 1.25) {
    \textbf{Override}\\Relabeling\\($\mathrm{score} > \theta$)
};

\begin{scope}[on background layer]
  \node[draw=blue!50, dashed, fill=blue!4, rounded corners=7pt,
        inner sep=14pt,
        fit=(detic)(clip)(override)(prompt),
        label={[font=\footnotesize\sffamily, text=blue!65, yshift=3pt]
               above:\textit{OvmmPerception}}
        ] (ovmm) {};
\end{scope}

\node[voxelbox] (voxel) at (11.4, 1.25) {
    \textbf{3D Voxel}\\World Map
};

\node[smbox] (sm) at (14.2, 0.5) {
    \textbf{State}\\Machine
};

\node[envbox] (env) at (16.8, 3.6) {Environment};

\node[actbox] (act) at (17.25, 0.709) {Act};

\draw[->] (img.east) -- ++(0.35,0) |- (detic.west)
    node[lbl, pos=0.72, above] {RGB-D};

\draw[->, densely dashed, color=blue!60]
    (img.east) -- ++(0.35,0) |- (clip.west)
    node[lbl, pos=0.72, below, color=blue!60] {RGB crops};

\draw[->] (detic.south) -- (clip.north)
    node[lbl, midway, right] {instance boxes};

\draw[->] (prompt.north) -- (clip.south);

\draw[->] (detic.east) -- ++(0,0) -| (override.north);

\draw[->] (clip.east) -- ++(0,0) -| (override.south);

\draw[->] (override.east) -- (voxel.west);

\draw[->] (voxel.south) |- ([yshift=-15pt]sm.west);

\draw[->] (ws.south) -- ++(0,-2.2) -| (sm.south);

\draw[->] (sm.north) |- (env.west);

\draw[->] ([yshift=6pt]sm.east) -- (act.west);

\draw[->] (act.north) -- (act.north |- env.south);

\draw[->] (env.north) -- ++(0,1.5) -| (dbox.north);

\end{tikzpicture}
}
\caption{Architecture of the modified HomeRobot agent. 
DETIC (green) handles spatial proposals and geometry; a frozen 
CLIP encoder (blue) serves as the sole object classifier through 
a threshold-gated override mechanism (red), making it the direct 
target of typographic attacks.}
\label{fig:arch}
\end{figure*}
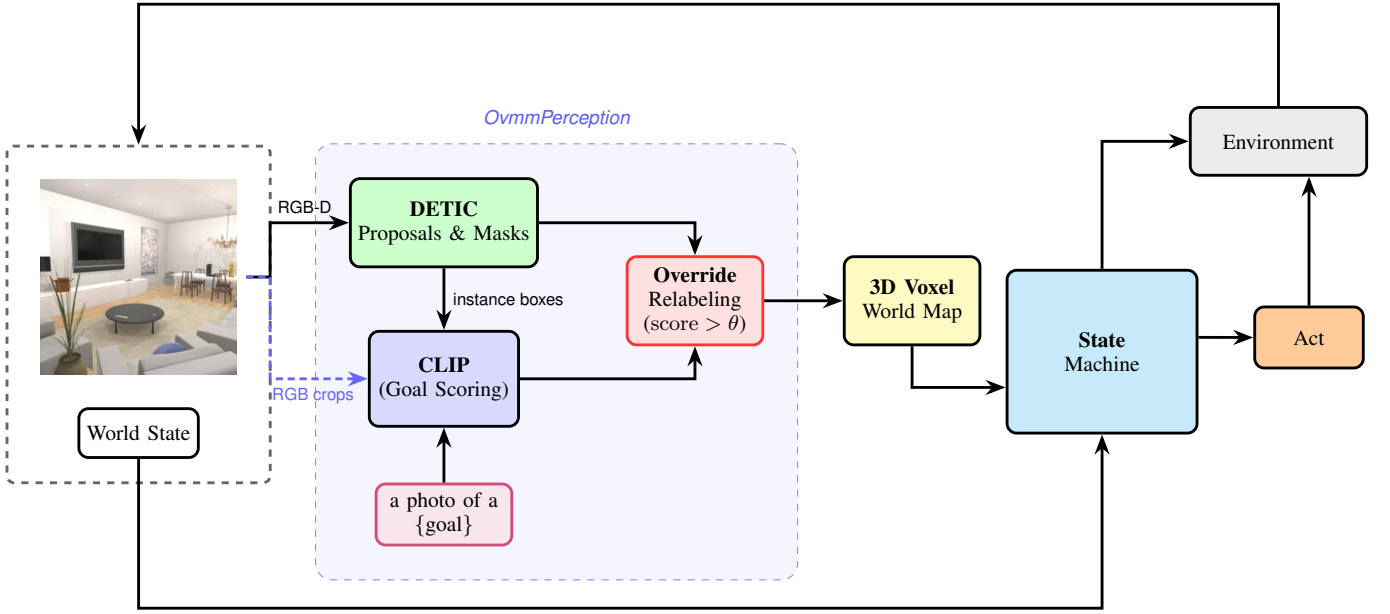

The resulting system implements a Sense-Plan-Act loop across three tightly coupled
components: a modified OvmmPerception module, a 3D Semantic Voxel Map, and a
heuristic finite-state machine. Fig.~\ref{fig:arch} illustrates how these
components interact and where the attack surface sits within the pipeline.

\textbf{OvmmPerception} serves as the system's perception front-end and constitutes
the primary attack surface. In the unmodified HomeRobot agent, a unified DETIC
component handles both spatial proposal generation and semantic labeling. Our
architecture decouples these two responsibilities: DETIC is retained exclusively
for spatial grounding, while a frozen CLIP encoder is introduced as the sole final
object identifier through a threshold-gated override mechanism. This separation
ensures that the effects of a typographic attack can be precisely localized to the
CLIP classification stage.

\textbf{DETIC} processes each RGB-D frame to produce per-instance bounding boxes,
segmentation masks, and depth-grounded coordinates. Its preliminary semantic 
labels are forwarded to the override stage but are discarded whenever CLIP's 
confidence exceeds the classification threshold $\theta = 0.28$. Critically, only 
the semantic label is overridden: the bounding box, segmentation mask, and 
depth coordinates produced by DETIC remain intact and are passed to the 
voxel map for backprojection and to the FMM planner for collision-free 
navigation, ensuring that geometric grounding is preserved regardless of 
whether an adversarial override occurs.

\textbf{Frozen CLIP encoder.} For each DETIC-detected instance, the frozen CLIP
encoder computes a text-image cosine similarity score between an RGB crop of that
instance and the natural-language goal prompt. If the highest-scoring instance
exceeds $\theta$, CLIP overrides DETIC's label and designates that instance as the
goal object. An adversary exploits this by attaching a typographic sticker whose
text aligns with the goal prompt to a distractor object, driving its similarity
score above $\theta$. Because the pipeline contains no input sanitization or
anomaly detection, the override is fully exploitable through crafted visual text
alone.

\textbf{3D Semantic Voxel Map.} The voxel map maintains persistent
spatial memory by backprojecting relabeled instance masks and depth
readings into a global voxel grid at 5\,cm resolution, as configured
in the HomeRobot agent~\cite{homerobot2023}. A misclassification
written into map state persists across timesteps: the agent may
continue navigating toward an adversarially labeled distractor even
after the sticker has left its field of view, amplifying the attack
beyond the adversary's window of visual influence.

From a security perspective, the voxel map functions as a spatial
cache that maps voxel coordinates $(x, y, z) \in \mathbb{R}^3$ to
semantic labels. A label is written on CLIP override and subsequently
read by the state machine and FMM planner on every downstream cycle
without re-verification. A successful typographic override therefore constitutes a cache
poisoning attack: a single malicious write at the Override Relabeling
stage (shown in red in Fig.~\ref{fig:arch}) corrupts the agent's
spatial memory, a form of semantic state corruption where the robot's
fundamental understanding of which object exists at a given location
is replaced with adversarially injected information. This drives all
subsequent navigation and manipulation decisions toward the wrong
object for the remainder of the episode, regardless of whether the
physical sticker remains visible. In several episodes, this persistence was directly
observable: the agent scored the adversarial sticker above
$\theta$ on first observation, writing the poisoned label
into the voxel map, then continued exploring the scene for
several seconds before returning to pick the sticker. The
pick was driven entirely by the cached map entry rather than
any fresh visual observation, confirming that a single
override event is sufficient to commit the agent to the
wrong object across subsequent planning cycles.

\textbf{Heuristic finite-state machine.} The state machine drives task execution
through four states: \textit{Navigate to Object},
\textit{Pick}, \textit{Navigate to Receptacle}, and
\textit{Place}. It queries the voxel map for the current target and dispatches
commands through two distinct paths. Navigation sub-goals are forwarded to the Act
module, where the Fast Marching Method (FMM) computes collision-free paths on a 2D
occupancy map and issues discrete \textit{Move} and \textit{Turn} actions to the
simulator. Manipulation actions (\textit{Pick} and \textit{Place}) bypass Act
entirely, issuing snap-based API calls directly to Habitat, which resolves them by
proximity check without modeling contact physics. A misclassification propagates
through both paths: the agent navigates to the wrong target via Act, then issues a
\textit{Pick} call for it, constituting the observable task failure used as the
primary attack outcome in our evaluation.

\section{Evaluation Setup}\label{sec:eval}

\subsection{Research Questions}

We structure our evaluation around five research questions
that together characterize the effectiveness, mechanism,
limits, and pipeline-level consequences of typographic
attacks in household robot manipulation.

\textbf{RQ1:} What does the full 1,199-episode validation set
consist of, and which episodes are attributable for attack
evaluation?

\textbf{RQ2:} Can a typographic sticker placed in a
household scene cause a capable manipulation agent to pick
the wrong object, and at what rate does this occur across
a controlled evaluation pool?

\textbf{RQ3:} Through what mechanism does a successful
attack operate: is the misclassification driven primarily
by the target sticker or by background text captured
incidentally in DETIC's bounding box crops?

\textbf{RQ4:} Under what conditions does the attack fail,
and can perceptual factors such as viewing angle,
occlusion, or crop visibility naturally bound its
reach without any defensive intervention?

\textbf{RQ5:} Does the attack shift the phase at which task
failure occurs, and can it drive the agent to mechanical
task completion with the wrong object?

\subsection{Attack Configuration}

Each episode $\tau \in \mathcal{E}^*$ is evaluated under a standardized
two-item configuration: the actual goal object $g_\tau$ is present in the scene
as usual, and one typographic sticker bearing the goal object name $g_\tau$
is placed on a flat surface along the agent's expected navigation path,
closer to the agent's starting position than the actual goal object.
This ensures the agent encounters and scores the sticker before ever
observing the legitimate target, giving the typographic cue the
first opportunity to poison the voxel map.

Sticker placement was designed to maximize visibility: stickers were
positioned on surfaces at heights consistent with the agent's camera
field of view, on unobstructed faces, and oriented toward the expected
approach direction. This placement strategy reflects a realistic threat
model in which an adversary with partial knowledge of the environment
can position a single typographic cue to intercept the agent's
perception pipeline prior to any observation of the legitimate target.

\textbf{Threshold selection.} The classification threshold $\theta$ is
set to 0.28. To determine this value, we inspected CLIP's cosine
similarity score distribution across all detected instances in a
representative set of episodes without any attack present. Non-goal
objects consistently scored between 0.17 and 0.25, with semantically
similar distractors approaching the upper end of this range. The true
goal object, when visible to the agent, scored between 0.28 and 0.30.
We therefore set $\theta = 0.28$ to sit above the distractor range
while remaining within the goal object's observed scoring range,
ensuring that only instances whose text-image similarity is consistent
with the goal prompt $g_\tau$ trigger an override of DETIC's label.
Raising $\theta$ above 0.28 begins excluding valid goal object
detections: at $\theta = 0.30$, no instance would ever exceed the
threshold and the override mechanism would never trigger, making goal
object identification impossible. Lowering $\theta$ below 0.28 pulls
non-goal objects into the override range: below 0.25, semantically
similar distractors begin triggering false overrides, and below 0.17
the threshold is so permissive that essentially any detected object
would be misclassified as the goal regardless of its visual or
semantic content. The two observed scoring ranges, 0.17 to 0.25 for non-goal objects
and 0.28 to 0.30 for goal objects, have zero overlap. $\theta = 0.28$ sits at the boundary of this gap, above the
distractor ceiling and within the goal object's range, making
it the optimal choice given the observed score distributions.
A system designer aware of typographic attacks cannot harden
the threshold by raising it above 0.28 without sacrificing
legitimate goal detection: since the true goal object scores
at most 0.30, any threshold above this range causes the agent
to fail its primary task entirely. Furthermore, an adversarial
sticker cannot be made to score arbitrarily higher than the
goal object, since the goal object label $g_\tau$ and the
sticker text are semantically identical, producing comparable
cosine similarity scores under the same CLIP encoder. The
vulnerability is therefore structural: the threshold that
enables open-vocabulary goal identification is the same
threshold that the attack exploits.

To characterize sensitivity to threshold selection, Table~\ref{tab:threshold}
summarizes the behavioral consequences of varying $\theta$ across the observed
score range. Because the non-goal score range ($0.17$-$0.25$) and goal object
score range ($0.28$-$0.30$) are non-overlapping, the viable range for $\theta$
is structurally constrained to the gap between them.

\begin{table}[h]
\caption{Behavioral consequences of varying the classification threshold
$\theta$ across the observed CLIP cosine similarity score ranges.}
\label{tab:threshold}
\begin{center}
\renewcommand{\arraystretch}{1.4}
\small
\begin{tabular}{>{\raggedright}p{0.9cm} >{\raggedright}p{3.0cm} >{\raggedright\arraybackslash}p{3.3cm}}
\toprule
\textbf{$\theta$} & \textbf{Override behavior} & \textbf{Consequence} \\
\midrule
$< 0.17$       & All detected objects trigger override       & Goal identification collapses; any object becomes goal \\
$0.17$-$0.25$ & Non-goal distractors trigger override       & Systematic false positives even without stickers \\
$0.28$         & Only goal-range instances trigger override  & Intended operating point \\
$0.30$         & Upper bound of goal object range            & Override barely fires; misses most goal detections \\
$> 0.30$       & No observed score exceeds threshold         & CLIP override never activates; goal identification impossible \\
\bottomrule
\end{tabular}
\end{center}
\end{table}

This sweep confirms that $\theta = 0.28$ is not an arbitrary choice but the
unique viable point given the observed score distributions. A system designer
cannot raise it above $0.28$ without losing valid goal detections, and cannot
lower it below $0.28$ without inducing false overrides on clean distractor
objects. The attack exploits this structural constraint: an adversarial sticker
bearing $g_\tau$ produces a similarity score in the same $0.28$-$0.30$ range
as the legitimate goal object, because both scores arise from CLIP comparing the
same text string $g_\tau$ against image evidence. No threshold value admits
legitimate goal detections while excluding adversarial ones.

\section{Results}\label{sec:results}

\subsection{RQ1: Episode Characterization and Attributability}

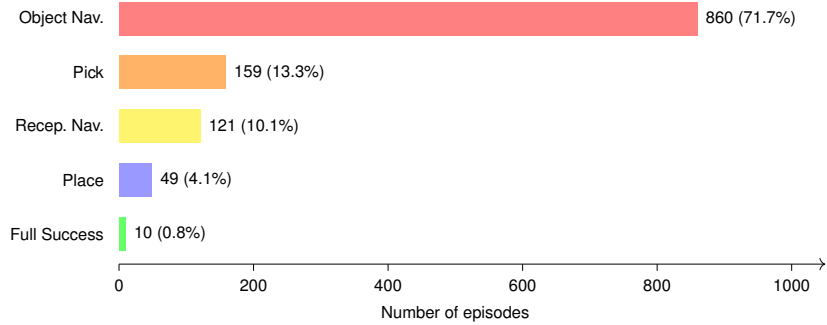
\begin{figure*}[!t]
\centering
\resizebox{0.6\textwidth}{!}{%
\begin{tikzpicture}
\tikzset{every node/.style={font=\small\sffamily}}

\def\xscale{0.01}  
\def\offset{2.5}
\def\bh{0.5}       
\def\sep{0.8}      

\def\ya{3.2}
\def\yb{2.4}
\def\yc{1.6}
\def\yd{0.8}
\def\ye{0.0}

\fill[red!50] (\offset,\ya) rectangle ({\offset+860*\xscale}, \ya+\bh);
\node[left] at (\offset-0.1, \ya+\bh/2) {\scriptsize Object Nav.};
\node[right] at ({\offset+860*\xscale}, \ya+\bh/2) {\scriptsize 860 (71.7\%)};

\fill[orange!60] (\offset,\yb) rectangle ({\offset+159*\xscale}, \yb+\bh);
\node[left] at (\offset-0.1, \yb+\bh/2) {\scriptsize Pick};
\node[right] at ({\offset+159*\xscale}, \yb+\bh/2) {\scriptsize 159 (13.3\%)};

\fill[yellow!70] (\offset,\yc) rectangle ({\offset+121*\xscale}, \yc+\bh);
\node[left] at (\offset-0.1, \yc+\bh/2) {\scriptsize Recep. Nav.};
\node[right] at ({\offset+121*\xscale}, \yc+\bh/2) {\scriptsize 121 (10.1\%)};

\fill[blue!40] (\offset,\yd) rectangle ({\offset+49*\xscale}, \yd+\bh);
\node[left] at (\offset-0.1, \yd+\bh/2) {\scriptsize Place};
\node[right] at ({\offset+49*\xscale}, \yd+\bh/2) {\scriptsize 49 (4.1\%)};

\fill[green!60] (\offset,\ye) rectangle ({\offset+10*\xscale}, \ye+\bh);
\node[left] at (\offset-0.1, \ye+\bh/2) {\scriptsize Full Success};
\node[right] at ({\offset+10*\xscale}, \ye+\bh/2) {\scriptsize 10 (0.8\%)};

\draw[->] (\offset,-0.2) -- ({\offset+1000*\xscale+0.5},-0.2);
\node[below] at ({\offset+500*\xscale}, -0.7) {\scriptsize Number of episodes};

\foreach \val in {0, 200, 400, 600, 800, 1000}{
    \draw ({\offset+\val*\xscale}, -0.2) -- ({\offset+\val*\xscale}, -0.3);
    \node[below] at ({\offset+\val*\xscale}, -0.3) {\scriptsize \val};
}

\end{tikzpicture}%
}
\caption{Pre-attack failure distribution across all 1,199 HomeRobot
validation episodes by phase of first failure.}
\label{fig:bar_failure}
\end{figure*}

The full HomeRobot validation set $\mathcal{E}$ consists of 1,199
episodes spanning diverse residential scenes with varying goal objects
and room configurations. We define an episode $\tau \in \mathcal{E}$
as \textit{fully successful}, $\tau \in \mathcal{E}^*_f$, if the
unattacked agent completes all four phases: object navigation, pick,
receptacle navigation, and place. We define $\tau$ as \textit{partially
successful}, $\tau \in \mathcal{E}^*_p$, if the agent completes object
navigation, pick, and receptacle navigation but fails only at the final
place step. All remaining episodes are filtered for attributability, as failures
in those cases cannot be unambiguously attributed to the adversarial
sticker. Evaluating on these episodes would produce false negatives
regarding the attack's perceptual effectiveness, since any observed
failure could reflect pre-existing agent limitations rather than the
typographic perturbation.
Our evaluation pool is then $\mathcal{E}^* = \mathcal{E}^*_f \cup
\mathcal{E}^*_p \subset \mathcal{E}$, where $|\mathcal{E}^*_f| = 10$,
$|\mathcal{E}^*_p| = 49$, and $|\mathcal{E}^*| = 59$.
Fig.~\ref{fig:bar_failure} shows the phase-level failure distribution
across all 1,199 episodes.

The dominant failure mode is \textit{Object Navigation},
accounting for 860 episodes (71.7\%). In these episodes the
agent fails to locate the goal object entirely, meaning no
pick attempt is ever made. Introducing a typographic sticker
into such an episode produces no attributable result: the
agent may still fail to navigate to the sticker for the same
perceptual and navigational reasons it failed to find the
goal object, making it impossible to isolate the sticker as
the cause of any observed failure. The 159 episodes (13.3\%) that fail at \textit{Pick} and the
121 episodes (10.1\%) that fail at \textit{Receptacle Navigation}
are excluded for a related but distinct reason. In these episodes
the agent may successfully locate and pick an object, but without
completing receptacle navigation and placement, the attack outcome
is incomplete: our goal is to demonstrate that a household robot
can be made to physically hand or deliver the wrong object to its
destination, not merely to pick it. An episode that terminates
before the full pick-and-place loop completes does not exhibit the
kinetic failure mode central to this work, where a misclassified
object travels through the entire pipeline and ends up in the wrong
place, an outcome that most directly reflects the real-world risk of
a robot handing a user the wrong, and potentially harmful, object.

The 49 episodes (4.1\%) that fail only at \textit{Place} are
retained. In these episodes the agent successfully navigates
to the goal object region, identifies and picks an object,
and navigates to the receptacle, failing only at the final
placement step. Because the agent has already demonstrated
it can complete every upstream phase, any change in what
object it picks under attack is directly attributable to the
typographic perturbation rather than to pre-existing agent
limitations. The 10 episodes (0.8\%) in which the agent
completes all four phases successfully are retained for the
same reason and represent the strongest attributability
condition.

The resulting pool of 59 episodes is the exhaustive set of HomeRobot validation
episodes satisfying the attributability requirement. It is not a sample drawn
from a larger population: every episode in $\mathcal{E}$ was inspected, and the
59 retained are the complete set where pre-existing agent failure cannot confound
the attack outcome. A larger pool could be produced by relaxing the
attributability criterion, but doing so would reintroduce confounds that make it
impossible to isolate the typographic perturbation as the cause of any observed
deviation. Within this pool the attack success rate is an exact figure rather than 
an estimate of a broader population parameter.

\subsection{RQ2: Attack Performance}

Of the 1,199 episodes in $\mathcal{E}$, 59 meet the attributability
criterion established in RQ1 and form the evaluation pool
$\mathcal{E}^*$ for this and all subsequent RQs. We evaluate
the attack across these 59 episodes using Attack Success Rate (ASR), defined as the fraction
of episodes in which the agent picks an object other than the goal
object $g_\tau$. Formally:

\begin{align}
\mathrm{ASR}(\mathcal{E}^*_f) &= \frac{7}{10} = 0.70 \\[4pt]
\mathrm{ASR}(\mathcal{E}^*_p) &= \frac{33}{49} \approx 0.673 \\[4pt]
\mathrm{ASR}(\mathcal{E}^*) &= \frac{40}{59} \approx 0.678
\end{align}

Table~\ref{tab:results} summarizes the attack outcomes across
$\mathcal{E}^*$. The attack achieves an overall ASR of 67.8\% (40/59). 
Among the 10 fully successful episodes $\mathcal{E}^*_f$, the ASR reaches
70.0\% (7/10), and among the 49 partially successful episodes
$\mathcal{E}^*_p$, the ASR is 67.3\% (33/49).

This result should be interpreted in the context of the evaluation conditions.
ASR and accuracy drop measure fundamentally different quantities: ASR is an
episode-level binary outcome spanning a full pick-and-place loop, while accuracy
drop is an image-level classification delta. No quantitative comparison between
them is meaningful, and we do not claim one. For orientation, SCAM~\cite{scam2025}
reports an average accuracy drop of 26 percentage points across 99 vision-language
models under real-world handwritten typographic attacks, with individual model
drops ranging from 2.67 to 66.37 percentage points; our 67.8\% episode-level ASR
reflects a structurally different and harder evaluation condition. Stickers are
placed physically in the simulated environment without perceptual optimization,
and the agent encounters them under uncontrolled viewing angles, distances, and
occlusion that no adversary can fully govern.
CHAI~\cite{chai2026} targets LVLM-driven embodied agents including autonomous
drones and ground vehicles, achieving up to 95.5\% ASR on aerial object tracking
and above 87\% in real-world robotic vehicle experiments by jointly optimizing
the semantic content and perceptual appearance of injected visual prompts. Both
SCAM and CHAI evaluate attacks under conditions that maximize adversarial text
exposure, either through standardized post-it note placement or jointly optimized
color, font, and positioning. Our evaluation imposes none of these advantages.

\begin{table}[htbp]
\caption{Attack success rate across evaluation episodes.}
\begin{center}
\renewcommand{\arraystretch}{1.3}
\small
\begin{tabular}{lccc}
\toprule
\textbf{Episodes} & \textbf{Total} & \textbf{Attacks} & \textbf{ASR} \\
\midrule
$\mathcal{E}^*_f$ (fully succ.)     & 10 & 7  & 70.0\% \\
$\mathcal{E}^*_p$ (partially succ.) & 49 & 33 & 67.3\% \\
\midrule
Combined $\mathcal{E}^*$            & 59 & 40 & 67.8\% \\
\bottomrule
\end{tabular}
\label{tab:results}
\end{center}
\end{table}

\subsection{RQ3: Attack Mechanism Breakdown}

To address RQ3 we analyze the 40 successful attacks from
$\mathcal{E}^*$. Of these, 37/40 (92.5\%) resulted in the agent
picking the primary malicious sticker directly, and 3/40 (7.5\%)
resulted in the agent picking the object immediately in front of
the malicious sticker. This breakdown confirms that the
attack operates primarily through CLIP's text-image similarity
mechanism rather than through incidental scene perturbation.

The 3 cases in which the agent picked the object in front of
the malicious sticker are attributable to the sticker being
partially occluded behind a distractor object. When DETIC generated
a bounding box crop for that distractor, the sticker was visible
in the background of the crop. CLIP then classified the crop
based on the background typographic sticker rather than the foreground
object, scoring it above $\theta = 0.28$ and causing the
agent to pick the foreground object as the goal object $g_\tau$. In all
three cases the foreground object was a knife. This reflects the
background classification failure mode documented in TidyBot, where
Wu et al. observe that CLIP classifies background objects as the
foreground target rather than restricting its response to the intended
subject of the crop, a consequence of CLIP performing classification
over the entire input region without spatial attention to object
boundaries. In our setting this failure mode is triggered not by a
background object but by a background typographic sticker, confirming
that the same mechanism extends to adversarially placed text.

To quantify the impact of map poisoning, we analyzed the agent's
behavior during the execution phase. In 100\% of successful attacks,
the agent proceeded to grasp the adversarial target based on
coordinates stored in the 3D Voxel Map. We observed that the agent
often classified the sticker early in the episode, sometimes wandering
the scene for several seconds before returning to the poisoned
coordinates. Although the sticker was visible in the robot's field of
view at the time of the pick, the agent issued the kinetic command
based solely on the cached semantic label. No secondary perceptual
check was performed to reconcile the typographic cue with the
underlying object geometry, confirming that a single poisoned write
to the spatial memory is sufficient to bypass all subsequent safety
margins in the loop.

\subsection{RQ4: Failure Attribution}

To address RQ4 we examine the 19 unsuccessful attack episodes
from $\mathcal{E}^*$. A successful attack in episode $\tau$ is
defined as any outcome in which the agent picks an object other
than the goal object $g_\tau$, including the malicious sticker
itself or an object immediately adjacent to the sticker. Of the
19 unsuccessful attack episodes, all produced behavior identical
to their pre-attack behavior. The 3 episodes from $\mathcal{E}^*_f$
resulted in correct goal object picks despite the sticker's presence,
and the 16 episodes from $\mathcal{E}^*_p$ failed at placement
exactly as in the unattacked agent, with no observable change
in navigation or classification behavior.

In all 19 cases, the adversarial sticker had no measurable effect
on the agent's classification decisions at any timestep, indicating
that the sticker's cosine similarity score never exceeded
$\theta = 0.28$. The threshold is the single gate through which
a typographic attack must pass; when perceptual conditions prevent
the sticker from being clearly captured within DETIC's bounding
box crop, CLIP receives insufficient visual evidence of the text
to drive the similarity score above $\theta$. Our findings show
that viewing angle, occlusion, and crop visibility are the primary
factors that naturally bound the attack's reach without any
defensive intervention, a result consistent with the perceptual
failure conditions independently identified in
CHAI~\cite{chai2026}. No mechanism beyond the physical visibility
constraints of the scene was responsible for the 19 failures,
confirming that the 40 successful attacks represent episodes where
those constraints were absent and CLIP's vulnerability to
typographic text was the direct and sole cause of task deviation.

\subsection{RQ5: Pre-Attack vs. Attack Failure Distribution}

The phase-level failure distribution shifts markedly once the
attack is introduced across $\mathcal{E}^*$. All 40 attack-induced
failures are concentrated at the \textit{Place} phase. We distinguish between two kinetic failure modes observed
across $\mathcal{E}^*$. The 7 successful attacks from
$\mathcal{E}^*_f$ constitute \textit{Hijacked Delivery}: the
agent completes the full Sense-Plan-Act loop, picks the
adversarial target, navigates to the receptacle, and
successfully places it there. These represent the most severe
outcome, where the attack drives the pipeline to mechanical
completion with the wrong object physically delivered to the
destination. The 33 successful attacks from $\mathcal{E}^*_p$
constitute \textit{Malicious Transport}: the agent picks the
adversarial target and physically carries it across the
household environment to the receptacle, where placement fails
as it did in the pre-attack condition. Although placement
failure was already present in these episodes, the typographic
sticker causes the agent to commit to and transport the wrong
object under adversarial direction, representing a persistent
kinetic deviation from the user's intent that is fully
attributable to the attack.

This phase-level contrast highlights the core contribution of
this work. Pre-attack failures across the full 1,199 episodes
are dominated by early-phase breakdowns, with 71.7\% of episodes
terminating before the agent locates the goal object, reflecting
fundamental navigational and perceptual limitations of the
unmodified agent. Our evaluation is deliberately scoped to
$\mathcal{E}^*$, the 59 episodes where these limitations are
absent, in order to isolate the specific failure mode introduced
by typographic attacks. Within this pool, the attack does not
change which phase fails but what causes the failure: a
misclassification committed at the perception stage propagates
with kinetic persistence through the voxel map, undetected across
navigation and manipulation phases, until the task terminates with
the wrong object. This propagation through the full Sense-Plan-Act
loop, from perceptual override to physical grasping of an
adversarial target, is precisely the kinetic failure mode that
no prior typographic attack benchmark, whether static or embodied,
has evaluated in a household manipulation context and that this
work is designed to expose.

\section{Discussion and Architectural Mitigations}

The vulnerability identified here stems from unconditional trust in perceptual
overrides that corrupt long-term spatial state. A hardened encoder alone is
insufficient: even a classifier that resists typographic attacks most of the time
will occasionally be overridden, and a single poisoned voxel map entry commits
the agent to the wrong object for the remainder of the episode. We therefore
propose three architectural mitigations that target the propagation pathway
rather than the classifier.

\textbf{Multi-View Voxel Confirmation} requires a semantic label to be
confirmed from $k$ distinct viewing angles or distances before it is committed
to persistent map state, preventing a single adversarial crop from poisoning the
voxel map.

\textbf{Geometric-Semantic Anomaly Detection} flags instances where a visual
crop contains high-contrast printed text via lightweight OCR and treats the
resulting joint-embedding output as an untrusted input requiring secondary
validation before any map write occurs.

\textbf{Grasp-Time Re-Query} re-runs the classifier against a fresh crop
immediately before issuing a pick action, rather than committing to the cached
map state. This breaks the persistence chain at the last actionable point before
kinetic commitment.

Each mitigation is implementable without modifying the vision-language encoder
and addresses the architectural condition that encoder-level defenses such as
Dyslexify~\cite{dyslexify2025} and
Defense-Prefix~\cite{defenseprefix2023} leave unresolved.

\section{Conclusion}\label{sec:conclusion}

This work establishes that typographic misclassification is a
real, measurable, and physically consequential threat in household
robot manipulation, achievable by a low-resource physical adversary
with nothing more than a printer and brief access to the environment.
A single sticker placed along the agent's navigation path is
sufficient to poison the 3D semantic map and drive the full
Sense-Plan-Act pipeline to deliver the wrong object to a target
receptacle, achieving a 67.8\% ASR under uncontrolled physical
conditions with no perceptual optimization.

These results carry implications beyond the specific architecture
evaluated here. The vulnerability demonstrated in this work is not unique to
CLIP: any vision-language encoder that aligns visual and
textual representations in a shared embedding space, including
SigLIP-based VLA models such as OpenVLA~\cite{openvla2024}
and LVLM-driven agents, inherits the same structural
susceptibility to typographic interference~\cite{scam2025}. As modular
manipulation pipelines increasingly integrate open-vocabulary
vision-language classifiers to enable natural language task
specification, the attack surface demonstrated here will grow
correspondingly. While encoder-level defenses such as Dyslexify~\cite{dyslexify2025} and
Defense-Prefix~\cite{defenseprefix2023} reduce misclassification rates at the
classifier stage, they do not address the architectural condition that makes the
manipulation setting uniquely dangerous. The preceding section proposes
three mitigations that target the propagation pathway directly. Treating VLA
classifier outputs as untrusted inputs requiring multi-stage validation before
kinetic commitment is a necessary architectural principle for securing the next
generation of open-vocabulary manipulation agents.

\section*{Acknowledgment}

The authors would like to thank the anonymous reviewers and the shepherd for
their valuable comments and suggestions.


\begin{thebibliography}{00}
    \bibitem{visualgenome2017} R. Krishna et al., ``Visual Genome: Connecting language and vision using crowdsourced dense image annotations,'' \textit{Int. J. Comput. Vis.}, vol. 123, no. 1, pp. 32--73, 2017.
    \bibitem{clip2021} A. Radford et al., ``Learning transferable visual models from natural language supervision,'' in \textit{Proc. ICML}, 2021.
    \bibitem{embclip2022} A. Khandelwal, L. Weihs, R. Mottaghi, and A. Kembhavi, ``Simple but effective: CLIP embeddings for embodied AI,'' in \textit{Proc. CVPR}, 2022, pp. 14809--14818.
    \bibitem{tidybot2023} J. Wu et al., ``TidyBot: Personalized robot assistance with large language models,'' \textit{Autonomous Robots}, 2023.
    \bibitem{cliport2021} M. Shridhar, L. Manuelli, and D. Fox, ``CLIPort: What and where pathways for robotic manipulation,'' in \textit{Proc. CoRL}, 2021.
    \bibitem{openvla2024} M. J. Kim et al., ``OpenVLA: An open-source vision-language-action model,'' arXiv:2406.09246, 2024.
    \bibitem{siglip2023} X. Zhai et al., ``Sigmoid loss for language image pre-training,'' in \textit{Proc. ICCV}, 2023.
    \bibitem{scam2025} J. Westerhoff et al., ``SCAM: A real-world typographic robustness evaluation for multimodal foundation models,'' arXiv:2504.04893, 2025.
    \bibitem{dyslexify2025} L. Hufe et al., ``Dyslexify: A mechanistic defense against typographic attacks in CLIP,'' arXiv preprint, 2025.
    \bibitem{goh2021} G. Goh et al., ``Multimodal neurons in artificial neural networks,'' \textit{Distill}, 2021. [Online]. Available: https://distill.pub/2021/multimodal-neurons
    \bibitem{scenetap2024} Y. Cao et al., ``SceneTAP: Scene-coherent typographic adversarial planner against vision-language models in real-world environments,'' arXiv:2412.00114, 2024.
    \bibitem{habitat2019} M. Savva et al., ``Habitat: A platform for embodied AI research,'' in \textit{Proc. ICCV}, 2019.
    \bibitem{robothor2020} M. Deitke et al., ``RoboTHOR: An open simulation-to-real embodied AI platform,'' in \textit{Proc. CVPR}, 2020.
    \bibitem{homerobot2023} A. Yenamandra et al., ``HomeRobot: Open-vocabulary mobile manipulation,'' in \textit{Proc. CoRL}, 2023.
    \bibitem{rt2_2023} B. Zitkovich et al., ``RT-2: Vision-language-action models transfer web knowledge to robotic control,'' in \textit{Proc. CoRL}, 2023.
    \bibitem{vlaattack2025} T. Wang et al., ``Exploring the adversarial vulnerabilities of vision-language-action models in robotics,'' in \textit{Proc. ICCV}, 2025.
    \bibitem{tex3d2025} J. Chen et al., ``Tex3D: Objects as attack surfaces via adversarial 3D textures for vision-language-action models,'' arXiv:2604.01618, 2025.
    \bibitem{freezevla2025} X. Wang et al., ``FreezeVLA: Action-freezing attacks against vision-language-action models,'' arXiv:2509.19870, 2025.
    \bibitem{chai2026} L. Burbano et al., ``CHAI: Command hijacking against embodied AI,'' in \textit{Proc. IEEE SaTML}, 2026.
    \bibitem{trojanrobot2024} X. Wang et al., ``Robot collapse: Supply chain backdoor attacks against VLM-based robotic manipulation,'' arXiv:2411.11683, 2024.
    \bibitem{loratk2024} H. Liu et al., ``LoRATK: LoRA once, backdoor everywhere in the share-and-play ecosystem,'' in \textit{Proc. EMNLP Findings}, 2025.
    \bibitem{defenseprefix2023} H. Azuma and Y. Matsui, ``Defense-Prefix for preventing typographic attacks on CLIP,'' in \textit{Proc. ICCVW}, 2023, pp. 3646--3655.
    \bibitem{detic2022} X. Zhou et al., ``Detecting twenty-thousand classes using image-level supervision,'' in \textit{Proc. ECCV}, 2022.
\end{thebibliography}
\end{document}